%% file: main.tex
\documentclass[%
 aip,
 amsmath,amssymb,floatfix,
 reprint,%
]{revtex4-1}

\usepackage{graphicx}
\usepackage{dcolumn}
\usepackage{bm}
\usepackage{xcolor}
\usepackage[normalem]{ulem}
\usepackage[hidelinks]{hyperref}
\usepackage[utf8]{inputenc}
\usepackage[T1]{fontenc}
\usepackage{mathptmx}
\usepackage{etoolbox}
\usepackage{cleveref}
\usepackage{booktabs}

\makeatletter
\def\@email#1#2{%
 \endgroup
 \patchcmd{\titleblock@produce}
  {\frontmatter@RRAPformat}
  {\frontmatter@RRAPformat{\produce@RRAP{*#1\href{mailto:#2}{#2}}}\frontmatter@RRAPformat}
  {}{}
}%
\makeatother
\begin{document}
\newcommand\redsout{\bgroup\markoverwith{\textcolor{red}{\rule[0.5ex]{2pt}{0.4pt}}}\ULon}

\newcommand{\fix}[1]{\textcolor{magenta}{ZNM: #1}}

\renewcommand{\sec}[1]{\section{#1}}
\newcommand{\ssec}[1]{\subsection{#1}}
\newcommand{\sssec}[1]{\subsubsection{#1}}

\input{macros}

\preprint{AIP/123-QED}
\title{Exchange-correlation entropy from the generalized thermal adiabatic connection}

\author{Brittany P. Harding}
\address{University of California, Merced, 5200 North Lake Road, Merced, CA 95343}
\author{Zachary Mauri}
\address{Stanford University, 450 Jane Stanford Way, Stanford, CA 94305}
\author{Vera Xie}
\address{University of California, Merced, 5200 North Lake Road, Merced, CA 95343}
\author{Aurora Pribram-Jones}
\email{apj@ucmerced.edu}
\address{University of California, Merced, 5200 North Lake Road, Merced, CA 95343}

\date{\today}

\begin{abstract}
Warm dense matter is a highly energetic phase characterized by strong correlations, thermal effects, and quantum mechanical electrons. Thermal density functional theory is commonly used in simulations of this challenging phase, driving the development of temperature-dependent approximations to the exchange-correlation free energy. 
Approaches using the adiabatic connection formula are well known at zero temperature and have been recently leveraged at non-zero temperatures as well.  In this work, a generalized thermal adiabatic connection (GTAC) formula is proposed, introducing a fictitious temperature parameter. This allows extraction of the exchange-correlation entropy $S\xc$ using simulated interaction strength scaling. This procedure, in turn, provides an avenue for approximating the exchange-correlation entropy in a way that preserves consistency with a chosen ground-state approximation. In addition, analysis of $S\xc$ as a function of interaction strength suggests new forms for approximations and GTAC itself offers a new framework for exploring both the exact and approximate interplay of temperature, density, and interaction strength across a wide range of conditions.
\end{abstract}


\maketitle

\sec{Introduction}


Warm dense matter (WDM) is a highly energetic phase that exists within the interiors of giant planets and in the atmospheres of white dwarf stars \cite{MHVT08,KRDM08,KDLM12}. WDM is also generated experimentally on the path to fusion ignition at flagship  facilities including Lawrence Livermore National Laboratory’s NIF and Sandia National Laboratory's Z Machine \cite{MSAA05,HCCD16,KD17,ZKHC22,KZCH22,AAAA22,MDDF23}. Simulations\cite{GMOB22,DMRT23,HBNF23} provide key information for both the design of WDM experiments and the analysis of their results, with DFT's balance of accuracy and efficiency driving its wide-spread use in various forms. Thermal density functional theory (DFT)\cite{HK64,M65,KS65} is commonly used to model this challenging phase and is considered the best practice for predictive WDM and high-energy density science (HEDS) calculations,\cite{SP88,KD09,GDRT14,RR14,VCW14,KDBL15,H17,JLWH23,WFGG17} often used to drive ab initio molecular dynamics (AIMD).

Accurate free-energy density functionals representative of the state conditions are necessary for reliable predictions of WDM properties\cite{KDT18} and for other systems in which temperature effects are important.\cite{GCG10} There are currently only a few approximate free-energy exchange-correlation (XC) functionals,\cite{KST12,KSDT14,KDT16,GDSM17,KDT18,KMH22,KPB23} and ground-state exchange-correlation functionals are employed with thermally weighted densities in almost all AIMD HEDS  simulations. Insights into the construction of accurate temperature-dependent functional approximations are at times informed by the growing list of temperature-dependent exact constraints and analytical tools, such as the adiabatic connection. 

The adiabatic connection\cite{HJ74,LP75,GL76} gives an exact expression for the ground-state exchange-correlation energy in terms of an integrand based on the fluctuation-dissipation theorem. The relationship described by the adiabatic connection formula (ACF) allows us to make a connection between the adiabatic connection and the exchange-correlation energy. Integration over coupling constant $\lambda$ from $\lambda=0$ to $\lambda=1$ smoothly connects the fictitious non-interacting reference system at $\lambda=0$ to the physical interacting system of interest at $\lambda=1$. 

The finite-temperature adiabatic connection formula (FTACF)\cite{PPFS11,PPGB14} introduces temperature dependence into the traditional AC integrand and was used to analyze an accurate, fully ab initio parameterization\cite{GDSM17} of the exchange-correlation free energy per electron at WDM conditions.\cite{HMP22} The generalized thermal adiabatic connection (GTAC) presented here generalizes the true FTACF to include variations in interaction strength and fictitious temperature, providing new pathways for constructing exchange-correlation approximations.

One of the most notorious complications of thermal DFT is the shift in interest from energy components to free energy components.  This shift is due to an entropic term, not only in the large-magnitude, non-interacting terms of the Kohn-Sham formalism, but also in the exchange-correlation. Though progress has been made in developing temperature-dependent exchange-correlation approximations, many DFT calculations still neglect the exchange-correlation entropy contribution in their practical calculations. For some researchers, this is due to an absence of exchange-correlation entropy approximations that are consistent with their use of ground-state approximations evaluated on thermal densities. Regardless of the root cause, examination of this component of the exchange-correlation free energy has established its importance.\cite{KCT16,SPB16,MSBV23,MSJB23} Formal analyses of the exchange-correlation entropy, as well as its relationship to the so-called zero-temperature approximation\cite{SPB16} and the often-substituted non-interacting entropy, can provide an avenue toward correction of this oversight in WDM simulations and beyond.

In this work, we will define the form of a generalized thermal adiabatic connection. The generalized form chosen allows variation of both interaction strength and a fictitious temperature, altering both the single-particle eigenstates and the occupations of those states, in addition to changing the coupling constant in a way consistent with defining this fictitious temperature as a thermal parameter. This GTAC approach is then demonstrated using known parametrizations of the exchange-correlation free energy per particle of the uniform gas, from which we extract approximations to the exchange-correlation entropy. We close with analysis of these interaction-strength-dependent $S\xc$ curves and discuss future directions for GTAC.

\sec{Background}

\ssec{Kohn-Sham DFT} 
Hohenberg and Kohn provided the basis for DFT,\cite{HK64} showing that if the exact ground-state density of a many-body interacting system is known, then the ground-state energy can be determined exactly:
    \ben\label{HK}
        E[n({\bf r})] \equiv \min_n \left\{\int d{\bf r}~ v({\bf r})n({\bf r}) + F[n({\bf r})]\right\},
    \een
where $v(\bf r)$ is the external potential, or system-dependent piece, and $F[n(\bf r)]$ is the universal functional, or system-independent piece consisting of the kinetic and electron-electron interaction energies:
    \ben\label{eq:uni}
        F[n(\br)] = T[n(\br)] + V\ee[n(\br)].
    \een

Kohn and Sham provided the framework for the practical implementation of DFT\cite{KS65}. The KS scheme begins with an imaginary system of non-interacting electrons that have the same density as the interacting problem. The electrons are embedded in a potential $v\s$ that causes these aloof, non-interacting electrons to imitate the true interacting system, driving a much more computationally efficient process for finding the ground-state density and corresponding ground-state energy. Since the electrons are non-interacting, the coordinates decouple and we may write the wavefunction as a product of single-particle orbitals satisfying
    \ben
        \left\{-\frac{1}{2}\nabla^2+v_s(\br)\right\}\phi_i(\br)=\varepsilon_i\phi_i(\br).
    \een
where $\varepsilon_i$ are the KS eigenvalues, $\phi_i$ are the corresponding KS orbitals yielding density $n(\br) = \sum_{i=1}^N |\phi_i(\br)|^2$, and $v\s$ is the KS potential (which is unique, by the HK theorem).
The KS Slater determinant is not considered an approximation to the true wavefunction but is a fundamental property of any electronic system uniquely determined by the electronic density.\cite{KS65,ABC} It is this non-interacting wavefunction, $\Phi$, that minimizes the kinetic energy to give the kinetic energy of the non-interacting electrons:
    \ben
        T\s[n]=\min_{\Phi\rightarrow n} \langle \Phi|\hatT|\Phi\rangle.
    \een
The subscript ``s'' will be used throughout this work to indicate non-interacting quantities. In terms of the non-interacting kinetic energy, Equation (\ref{eq:uni}) becomes
    \ben
        F[n(\br)] = T\s[n(\br)] + V\ee[n(\br)] + U\H[n(\br)] + E\xc[n(\br)],
    \een
where $U\H$ is the classical electrostatic repulsion, or Hartree energy, and $E\xc$ is the exchange-correlation (XC) energy. The exchange-correlation energy collects the remainder of interactions not captured by the Hartree energy, such as the kinetic correlation, 
    \ben
        T\c = T - T\s,
    \een
or the difference between the exact kinetic energy and the non-interacting kinetic energy. \KS DFT is exact as long as we have the exact exchange-correlation functional for arbitrary physical systems of interest. Unfortunately, this is not the case for the vast majority of systems, and approximations must be employed for practical calculations.


\ssec{Thermal DFT}
\label{sec:FTDFT}

DFT can be extended to finite temperatures by generalizing the Hohenberg-Kohn theorems and Kohn-Sham equations to thermal systems. Working within the grand canonical ensemble, Mermin generalized the HK theorems to equilibrium systems at finite temperatures constrained to fixed temperature and chemical potential\cite{M65}. The grand canonical potential of the system is written
    \ben
        \hat{\Omega}=\hatH-\tau\hat{S}-\mu\hat{N},
    \een
where $\hatH$ is the ground-state Hamiltonian, $\tau$ is the temperature, $\mu$ is the chemical potential, $\hat{N}$ is the particle-number operator, and $\hat{S}$ is the entropy operator,
    \ben
        \hat{S} = -k_B \, \mathrm{ln} \, \hat{\Gamma}.
    \een
The statistical operator $\hat{\Gamma}$ is
    \ben
        \hat{\Gamma} = \sum_{N,i} w_{\scriptscriptstyle\rm N,i} |\Psi_{\scriptscriptstyle\rm N,i}\rangle\langle \Psi_{\scriptscriptstyle\rm N,i}|,
    \een
where $|\Psi_{\scriptscriptstyle\rm N,i}\rangle$ are the orthonormal $N$-particle states, and $w_{\scriptscriptstyle\rm N,i}$ are the normalized statistical weights satisfying $\sum_{\scriptscriptstyle\rm N,i}w_{\scriptscriptstyle\rm N,i}=1$. The statistical operator yields the thermally weighted, equilibrium density. The Mermin-Kohn-Sham (MKS) equations,\cite{M65,KS65} which resemble the ground-state KS equations, but with temperature-dependent eigenstates, eigenvalues, and effective potential, yield the MKS density,

    \ben
        n\t({\bf r})=\sum_i f_i\t |\phi\t_i(\bf r)|^2,
    \een
with $\phi_i^\tau({\bf r})$ equal to the $i^{\rm th}$ eigenstate and $f_i^\tau$ equal to the state's corresponding Fermi occupation. The MKS density is equal to the exact equilibrium density at temperature $\tau$ by definition.

To define the exchange-correlation free energy, we decompose the universal functional and write the free energy as,\cite{PPGB13}
    \ben
        A=T\s-\tau S\s+U+A\xc\t+V\ext.
    \een
Here, $T\s$ is the non-interacting kinetic energy, $S\s$ is the non-interacting entropy, $U$ is the classical electrostatic repulsion, and $V\ext$ is the external potential\cite{FW71}. The temperature-dependent exchange-correlation free energy by definition is
    \ben
        A\xc\t=\big(T-T\s\big)-\tau\big(S-S\s\big)+\big(V\ee-U\H\big),
    \een
where $T[n]$ and $S[n]$ are the interacting kinetic energy and entropy, and $V_{ee}[n]$ is the electron-electron interaction energy.

\sec{Results}
In this section, we propose a generalized thermal adiabatic connection and show how to use this form to extract expressions for $S\xc$. 

\ssec{The Generalized Thermal Adiabatic Connection}
We begin by introducing a generalized thermal adiabatic connection using a somewhat familiar\cite{Y98} generalized electron-electron interaction operator,

    \bea
        \hatV\ee(\br^2,\lambda,\tilde\tau)=\sum_{i\neq j}\nu (|\br_i-\br_j|,\lambda,\tilde\tau),
    \eea
where $\lambda$ is a scaling factor similar to the linear scaling factor in the traditional adiabatic connection formula and $\tilde{\tau}$ is a parameter taking the form of a fictitious temperature. In contrast to other choices one might make, here we are using a form that mimics the original FT adiabatic connection formula, connecting the non-interacting and exact systems. Here, we generalize to allow thermal-like occcupations at smoothly varying ``temperatures.'' In some ways, this appears to mimic the zero-temperature generalized KS framework for hybrid functionals\cite{GNGK20}, in that there are two different scaling parameters, but our choice here maintains the KS framework for all values of $\tilde{\tau}$ and $\lambda$, allowing for the smooth and gradual scaling of the interaction strength and the associated minimizing ensemble. Scaling conditions present in the true FT system are maintained, as described in Section \ref{sec:SimScale}. As will become obvious below, this formalism can also be described outside of a GKS-like framework, by straightforward application of the fundamental theorem of calculus. In this way, we simultaneously connect the zero-temperature system and the truly thermalized system across the range of non-interacting to fully interacting electronic systems, via a $\tilde{\tau}$- and $\lambda$-dependent effective potential.

Following the usual path for the FT ACF, we can then define the universal functional, 
 
    \ben \label{min}
        \begin{split}
            F^{\tilde\tau,\lambda}[n,\nu^{\tilde\tau,\lambda}]&= F[n,\nu (|\br_i-\br_j|,\lambda,\tilde\tau)] \\
            &=\min_{\Gamma\rightarrow n}\text{Tr}~\hat{\Gamma}\left\{\hatT-\tilde\tau\hat{S}+\hatV\ee(\br^N,\lambda,\tilde\tau)\right\} \\
            &=\sum\N\sum_{i} w_{N,i} \langle \Psi_{N,i}|\hatT-\tilde\tau\hat{S}+\hatV\ee(\br^N,\lambda,\tilde\tau)|\Psi_{N,i}\rangle.
        \end{split}
    \een

This form of the electron-electron interaction not only alters the strength of interaction via the coupling constant, but also gives the interaction potential functional more flexibility by letting it fluctuate away from the physical temperature. All of these fluctuations are captured in the exchange-correlation free energy, similar to those generated by the coupling constant alone in the traditional adiabatic connection.

The effective potential above coincides with the true KS potential for the system of interest only at $\tilde{\tau}=\tau$ and $\lambda=1$. To ensure this, we can explicitly define our effective potential at specific interaction strengths:

    \bea
        \nu \big(|\br_i-\br_j|,\lambda=0,\tilde\tau\big)=0,
    \eea

\noindent and

    \bea
        \nu \big(|\br_i-\br_j|,\lambda=1,\tilde\tau\big)=\frac{1}{|\br_i-\br_j|}.
    \eea

\noindent In this way, we can write the universal functional in terms of non-interacting KS quantities,

    \ben \label{KS}
        \begin{split}
            F[n,\nu\big(|\br_i-\br_j|,\lambda=0,\tilde\tau\big)]&=F^{\tilde\tau,\lambda=0}[n,\nu^{\tilde\tau,\lambda}] \\
            &=T\s[n]-\tilde\tau S\s[n] \\
            &=K\s^{\tilde\tau}[n]
        \end{split}
    \een

\noindent and in terms of the Mermin functional,

    \ben \label{Mermin}
        \begin{split}
            F[n,\nu\big(|\br_i-\br_j|,\lambda=1,\tilde\tau\big)]&=F^{\tilde\tau,\lambda=1}[n,\nu^{\tilde\tau,\lambda}] \\
            &=K^{\tilde\tau}[n]+V\ee[n] \\
        \end{split}    
    \een

\noindent Now, defining the exchange-correlation free energy in terms of (\ref{KS}) and (\ref{Mermin}),

    \ben
        \begin{split}
            A\xc\t[n]&=\left(K\t[n]+V\ee[n]\right)-\left(K\s\t[n]+U\H[n]\right) \\
            &=F\t[n]-F^{\tau,\lambda=0}[n]-U\H[n] \\
            &=\int_0^1 d\lambda \frac{dF^{\tau,\lambda}[n,\nu^{\tau,\lambda}]}{d\lambda}-U\H[n].
        \end{split}
    \een

\noindent Adding and subtracting $F^{\tilde\tau=0,\lambda}[n,\nu^{\tau,\lambda}]$,

        \begin{multline*}
            A\xc\t[n]=\int_0^1 d\lambda \biggr\{ 
            \biggr(\int_0\t d\tilde\tau\frac{d^2F^{\tilde\tau,\lambda}[n,\nu^{\tilde\tau,\lambda}]}{d\lambda d\tilde\tau}\biggr)\\
            +F^{\tilde\tau=0,\lambda}[n,\nu^{\tilde\tau,\lambda}]\biggr\}-U\H[n],
        \end{multline*}
and pulling the $\lambda$ derivative outside of the temperature integral,        
        \begin{multline*}
            A\xc\t[n]=\int_0^1 d\lambda \frac{d}{d\lambda} \biggr\{\biggr(\int_0\t d\tilde\tau \frac{dF^{\tilde\tau,\lambda}[n,\nu^{\tilde\tau,\lambda}]}{d\tilde\tau}\biggr|_{\lambda}\biggr) \\+ F^{\tilde\tau=0,\lambda}[n,\nu^{\tilde\tau,\lambda}]\biggr\}-U\H[n].
        \end{multline*}

\noindent Since the density is held fixed, $\int_0^1 d\lambda U\H[n]=U\H[n]$, and we find

        \begin{multline*}
             A\xc\t[n]=\int_0^1 d\lambda \frac{d}{d\lambda} \biggr\{\biggr(\int_0\t d\tilde\tau ~\sum\N\sum_i w_{N,i}\\
             \biggr\langle \Psi_{N,i}^{\tilde\tau,\lambda}\biggr|\frac{\partial \hat{F}^{\tilde\tau,\lambda}[n,\nu^{\tilde\tau,\lambda}]}{\partial\tilde\tau}\biggr|\Psi_{N,i}^{\tilde\tau,\lambda}\biggr\rangle\biggr)
             -U\H[n]\biggr\}\\
             + \int_0^1 d\lambda \frac{U\xc^{\tau=0,\lambda}[n]}{\lambda},
        \end{multline*}

\noindent where the last term is the ground-state adiabatic connection. Now rewriting $\frac{\partial\hat{F}^{\tilde\tau,\lambda}[n,\nu^{\tilde\tau,\lambda}]}{\partial\tilde\tau}$,

        \begin{multline*}
             A\xc\t[n]=\int_0^1 d\lambda \frac{d}{d\lambda} \biggr\{\biggr(\int_0\t d\tilde\tau ~\sum_{N,i} w_{N,i}\biggr\langle \Psi_{N,i}^{\tilde\tau,\lambda}\biggr|-\hat{S}\\
             +\frac{\partial \hatV\ee(\br\N,\lambda,\tilde\tau)}{\partial\tilde\tau}\biggr|\Psi_{N,i}^{\tilde\tau,\lambda}\biggr\rangle\biggr)-U\H[n]\biggr\}\\
             + \int_0^1 d\lambda \frac{U\xc^{\tau=0,\lambda}[n]}{\lambda}.
        \end{multline*}


\noindent Since the eigenstates $\Psi^{\tilde\tau,\lambda}_{N,i}$ are variational extrema, $\frac{d\Psi^{\tilde\tau,\lambda}_{N,i}}{d\lambda}$ contributions vanish. This implicit dependence on $\lambda$ is the only $\lambda$-dependence of $S^{\tau,\lambda}[n]$. Thus, $\frac{dS^{\tau,\lambda}[n]}{d\lambda}$ also vanishes, and we have

        \begin{multline*}
             A\xc\t[n]=\int_0^1 d\lambda \int_0\t d\tilde\tau ~\frac{d}{d\lambda}~ \sum_{N,i} w_{N,i} \biggr\langle \Psi_{N,i}^{\tilde\tau,\lambda}\biggr|\frac{\partial \hatV\ee(\br\N,\lambda,\tilde\tau)}{\partial\tilde\tau}\biggr|\Psi_{N,i}^{\tilde\tau,\lambda}\biggr\rangle\\
             -\int_0^1 d\lambda \frac{dU\H[n]}{d\lambda} + \int_0^1 d\lambda \frac{U\xc^{\tau=0,\lambda}[n]}{\lambda}
        \end{multline*}

Identifying the final term as the zero-temperacture exchange-correlation energy expressed in the traditional ACF, we can now write the exchange-correlation free energy in the GTAC formalism, here with the ground-state exchange-correlation {\it evaluated on the finite-temperature density} as its reference point:
        \bea\label{end}
            A\xc\t[n]= E\xc[n]+\int_0^1 d\lambda \int_0^{\tau} d\tilde\tau \frac{\partial}{\partial\tilde\tau}\frac{U\xc^{\tilde\tau,\lambda}[n]}{\lambda}.
        \eea


\ssec{Extraction of Exchange-Correlation Entropy}

    \begin{figure}[htbp]
        \centering
        \includegraphics[width=\columnwidth]{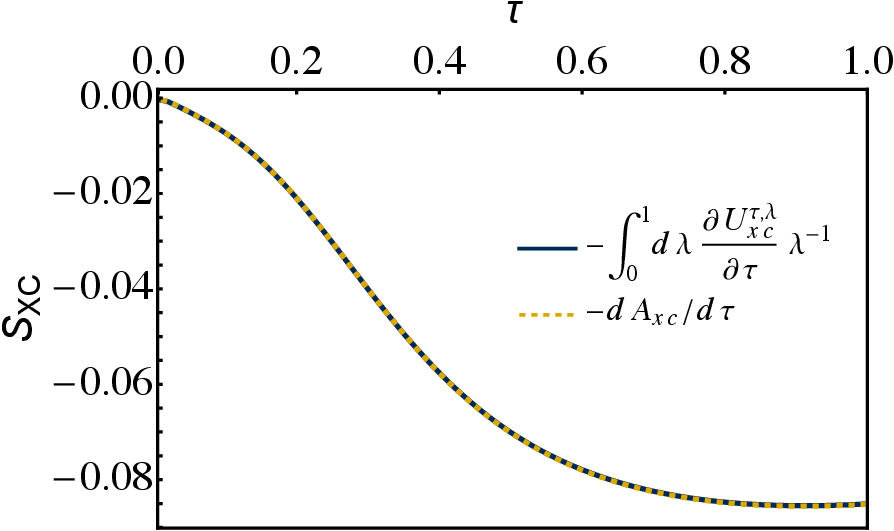}
        \caption{The exchange-correlation entropy obtained from expressions (\ref{eq:Maxwell}) and (\ref{eq:Sxc}) for the GDSM17 parameterization and compared for $r\s=1$ and $\tau\rightarrow1$.}
        \label{fig:dAxc}
    \end{figure}

The FTAC\cite{HMP22} can be rewritten as a partial differential of the exchange-correlation free energy with respect to $\lambda$, and again rewritten in terms of an integral over fictitious temperature:
    \ben
        A\xc\t[n]=\int_0^1 d\lambda \frac{\partial A\xc\t[n]}{\partial\lambda}=\int_0^1 d\lambda\frac{\partial}{\partial\lambda}\int_0\t d\tilde\tau \frac{\partial A\xc^{\tilde\tau}[n]}{\partial\tilde\tau}.
    \een
Via Leibniz's rule, we can switch the ordering of the integral and derivative for this definite integral:
    \ben
        \int_0^1 d\lambda \int_0\t d\tilde\tau \frac{\partial^2 A\xc^{\tilde\tau}[n]}{\partial\lambda \partial\tilde\tau}.
    \een
We thus obtain an object that is a mixed partial differential with respect to temperature and interaction strength. The explicit temperature dependence of this object is unknown for real, arbitrary systems outside of limiting conditions, but we can take advantage of the fact that the mixed partial derivatives are symmetric by virtue of assumed continuity, which must be true for the mixed partial in (\ref{eq:mixed}) to be integrable: 
    \ben\label{eq:mixed}
    \frac{\partial^2A\xc\t[n]}{\partial\tau\partial\lambda}=\frac{\partial^2A\xc\t[n]}{\partial\lambda\partial\tau}.
    \een
\noindent This symmetry is mirrored by Maxwell-style equations,\cite{BSGP16}

    \ben\label{eq:Maxwell}
        \left(\frac{\partial U\xc[n]}{\partial \tau}\right)_\lambda = -\lambda\left(\frac{\partial S\xc[n]}{\partial \lambda}\right)_\tau,
    \een
which relate the temperature-dependent exchange-correlation potential to the lambda-dependent exchange-correlation entropy. $U\xc^{\tau,\lambda}$, or the potential XC, comes from the FTAC integrand (Equation \ref{eq:Wxc}): $W\xc^{\tau,\lambda}=U\xc^{\tau,\lambda}/\lambda$. The exchange-correlation entropy can be extracted by taking the temperature derivative of $U\xc^{\tau,\lambda}$ and integrating over $\lambda$ for a given temperature $\tau$,
    \ben
        S\xc\t[n] = -\int_0^1 d\lambda ~\frac{1}{\lambda} \frac{\partial U\xc^{\tau,\lambda}[n]}{\partial\tau}.
    \een
This was done numerically for the corrected KSDT14\cite{KSDT14} and GDSM17\cite{GDSM17} parameterizations of the exchange-correlation free energy of the UEG at WDM conditions. Fig. \ref{fig:dAxc}
shows the exchange-correlation entropy obtained for the GDSM17 parameterization for an $r\s$ value of $1$ as temperature $\tau$ varies from $0$ to $1$. Also shown in Fig. \ref{fig:dAxc} is the exchange-correlation entropy obtained from the more familiar thermodynamic relationship,
    \ben\label{eq:Sxc}
        S\xc\t[n] = -\frac{\partial A\xc\t[n]}{\partial \tau}.
    \een
This comparison serves as verification that our GTAC extraction process using simulated interaction strength scaling and numerical integration yields the correct $S\xc$ for the FT UEG. In this way, we can be confident that the features we see in the next section are the result of the mathematical objects and not merely artifacts of our extraction technique.

\ssec{Numerical Demonstrations}
Tied coordinate—temperature—interaction strength scaling was applied to two well-known parameterized functions of the exchange-correlation free energy of the UEG at WDM conditions\cite{KSDT14,KDT18,GDSM17} to construct the proper form of the FTAC integrand. Temperature and interaction strength effects were then obtained by allowing the FTAC integrand to vary over a range of $\tau$ and $\lambda$, creating a 3D object that varies in both tempeature and interaction strength. Numerical differentiation was performed on the data to obtain the temperature dependence of the FTAC integrand extracted from each parameterization. Next, numerical integration was applied to obtain the exchange-correlation entropy for a given temperature and value of $r\s$. All plots and numerical work were done using Mathematica version 13.3; sample notebooks are available upon request.

\subsubsection*{Simulated Interaction Strength Scaling}
\label{sec:SimScale}
While quantum mechanical operators scale in a simple manner, the scaling of density functionals is not as straightforward. A useful relationship in DFT that is commonly taken advantage of is the relationship between coordinate scaling and interaction strength scaling.\cite{LP85,ABC} When the length scale of the system is altered by a factor of $\gamma$, the density of the system is expressed as,
    \ben
n_\gamma(\textbf{r})=\gamma^3 n(\gamma\textbf{r})
    \een
where the scaling factor out front preserves normalization of the system. This scaling of the density causes the non-interacting functionals into which they are input to exhibit power law scaling at zero temperature.  At finite temperature, these non-interacting functionals require scaling of the temperature to maintain these scaling relationships. At both zero and finite temperatures, introducing electron-electron interaction into the picture, such as when examining functionals with correlation components, one must also scale the interaction strength to achieve such neat and tidy power law scaling. This leads to the aforementioned tied coordinate-temperature-interaction strength scaling within thermal DFT.

In the case of the UEG, our focus in this work, interaction strength scaling of the density functions describing the system can be expressed in terms of temperature and density scaling, $\gamma\rightarrow\frac{1}{\lambda}$. For example, the exchange-correlation free energy at interaction strength $\lambda$ and temperature $\tau$ is written in terms of the same function, evaluated at a scaled density and scaled temperature, scaled quadratically overall:
    \ben
        a\xc^{\scriptscriptstyle\rm\tau,\lambda}(n)=\lambda^2a\xc^{\tau/\lambda^2}\left(n_{1/\lambda}\right).
    \een
The exchange free energy, expressed below in terms of the Wigner-Seitz radius, $r\s$, can be extracted from the exchange-correlation free energy by scaling to the high-density limit of the thermal UEG,
    \ben
        a_{\scriptscriptstyle\rm X}^{\tau}(r\s)=\lim_{\gamma\to\infty}\frac{a\xc^{\scriptscriptstyle\rm\gamma^2\tau}\left(\frac{r\s}{\gamma}\right)}{\gamma}.
    \een
The correlation free energy per particle is obtained by subtracting off the exchange component of the exchange-correlation free energy per particle,
    \ben
        a_{\scriptscriptstyle\rm C}\t(r\s)=a\xc\t(r\s)-a_{\scriptscriptstyle\rm X}\t(r\s).
    \een
Next, we construct the correlation component of the FTAC integrand by extracting the potential contribution from the correlation free energy per particle,
    \ben
        u\t_{\scriptscriptstyle\rm C}\left(\frac{r\s}{\gamma}\right)=-\gamma\frac{\partial a_{\scriptscriptstyle\rm C}^{\rm\gamma^2\tau}\left(\frac{r\s}{\gamma}\right)}{\partial\gamma}+2a_{\scriptscriptstyle\rm C}^{\rm\gamma^2\tau}\left(\frac{r\s}{\gamma}\right).
    \een
Using the definition of the FTAC integrand for the uniform gas in terms of the potential contribution, $u_c$, and the relationship for extracting $a_x$, we obtain the FTAC integrand,
    \ben\label{eq:Wxc}
        W^{\tau,\lambda}\xc(r\s)=\lambda u_{\scriptscriptstyle\rm C}^{\tau/\lambda^2}(\lambda r\s) + \lim_{\gamma\to\infty}\frac{a\xc^{\gamma^2\tau}(r\s/\gamma)}{\gamma}.
    \een

 \begin{figure}[htbp]
            \centering
            \includegraphics[width=\columnwidth]{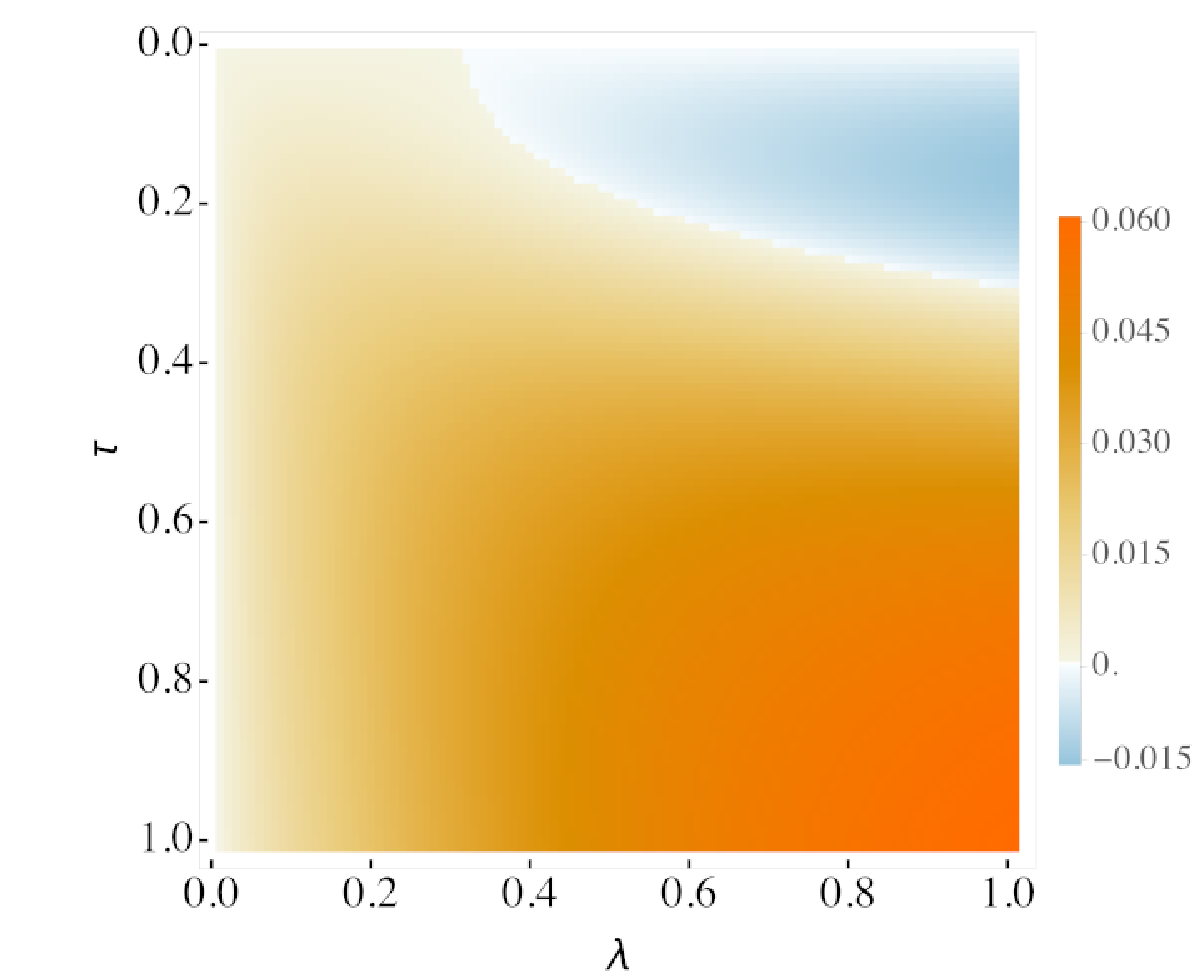}
            \caption{The temperature derivative of the potential exchange-correlation obtained from the corrected KSDT14 parameterization\cite{KDT18} is shown for $r\s=1$, $0\leq\tau\leq 1$ and $0\leq\lambda\leq1$.}
            \label{fig:2DCorrKSDT1}
       \end{figure}   

\subsubsection*{Results and Discussion}
The partial derivative of the potential exchange-correlation with respect to temperature was obtained numerically and is shown for two well-known paramaterizations of the XC free energy of the warm, dense UEG in Figs. \ref{fig:2DCorrKSDT1} and \ref{fig:2DGDSM1}. The 2D plots provide details on how the potential exchange-correlation varies with temperature. 

The blue regions in both figures indicate negative values of the temperature derivative and appear localized around temperatures below one third of the Fermi energy and interaction strengths above one third of the full interaction strength. This indicates that the potential piece of the exchange-correlation decreases with respect to temperature in low-temperature regimes of partially to fully interacting electrons. For the parameters tested, the temperature derivative remains positive regardless of the temperature for non-interacting and partially interacting ($\lambda\leq 0.3$) electrons.       

Fig. \ref{fig:2DGDSM1} shows the temperature derivative of the potential exchange-correlation for an $r\s$ value of $1$ obtained from the highly optimized GDSM17 parameterization. 
       \begin{figure}[htp]
            \centering
            \includegraphics[width=\columnwidth]{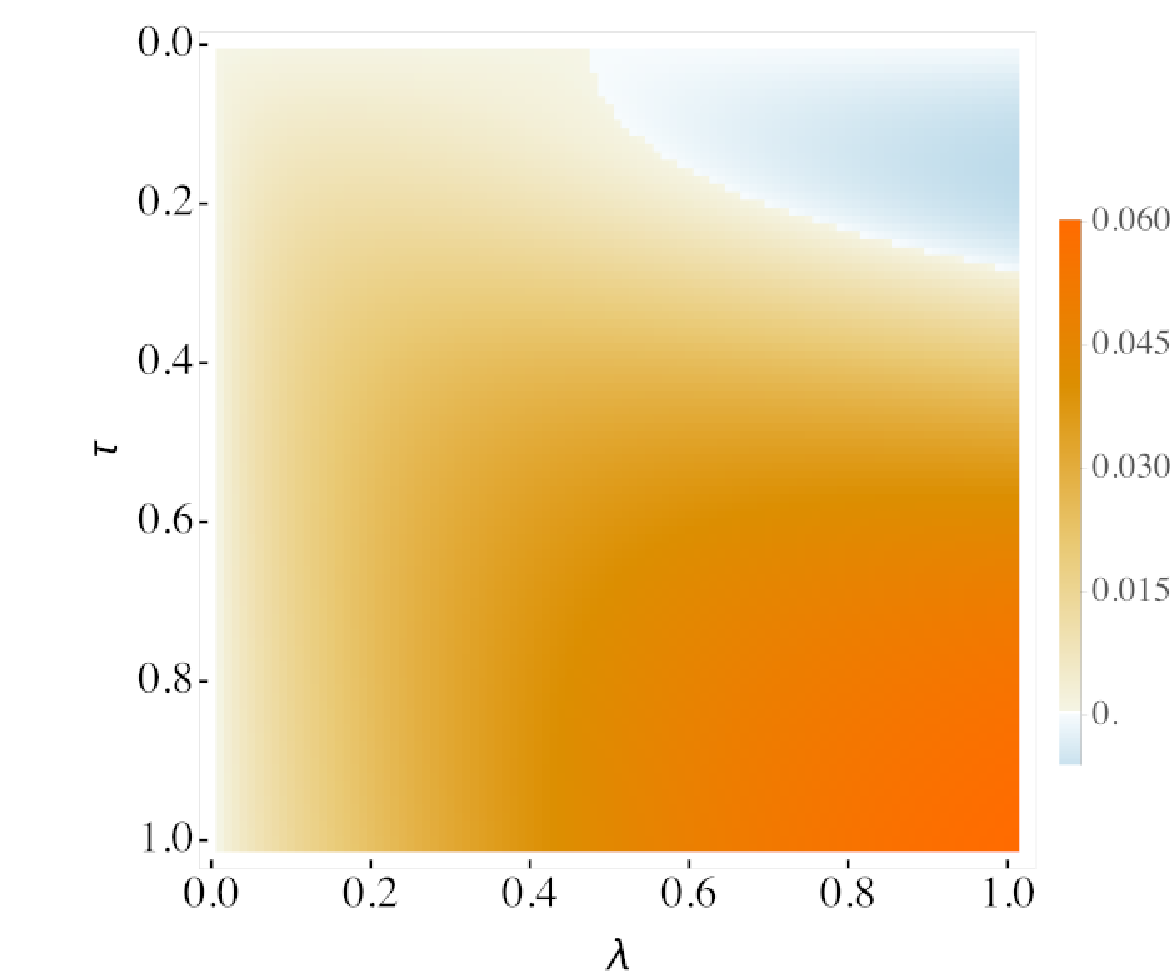}
            \caption{The temperature derivative of the potential exchange-correlation obtained from the GDSM17 parameterization\cite{GDSM17} is shown for $r\s=1$, $0\leq\tau\leq 1$ and $0\leq\lambda\leq1$.}
            \label{fig:2DGDSM1}
       \end{figure}
The negative values remain in the $\tau\leq 0.3$ region, but appear more localized at interaction strengths above $0.5$, or half, the full interaction strength compared to the temperature derivative plotted in Fig. \ref{fig:2DCorrKSDT1}.

Since $V_{ee}=U\H +U\xc$ and $U\xc\leq0$,\cite{ABC,LP93} when $U\xc< 0$,

    \begin{align}
        U\xc &= V\ee - U\H\\
        0 &\ge V\ee - U\H\\
        V\ee &\leq U\H.          
    \end{align}
Since both $V\ee$ and $U\H$ are both repulsive terms and thus, positive in sign, the electron-electron interaction energy is always larger in magnitude than the full Coulombic repulsion in our examples.      

\begin{figure}[htbp]
    \centering
\includegraphics[width=\columnwidth]{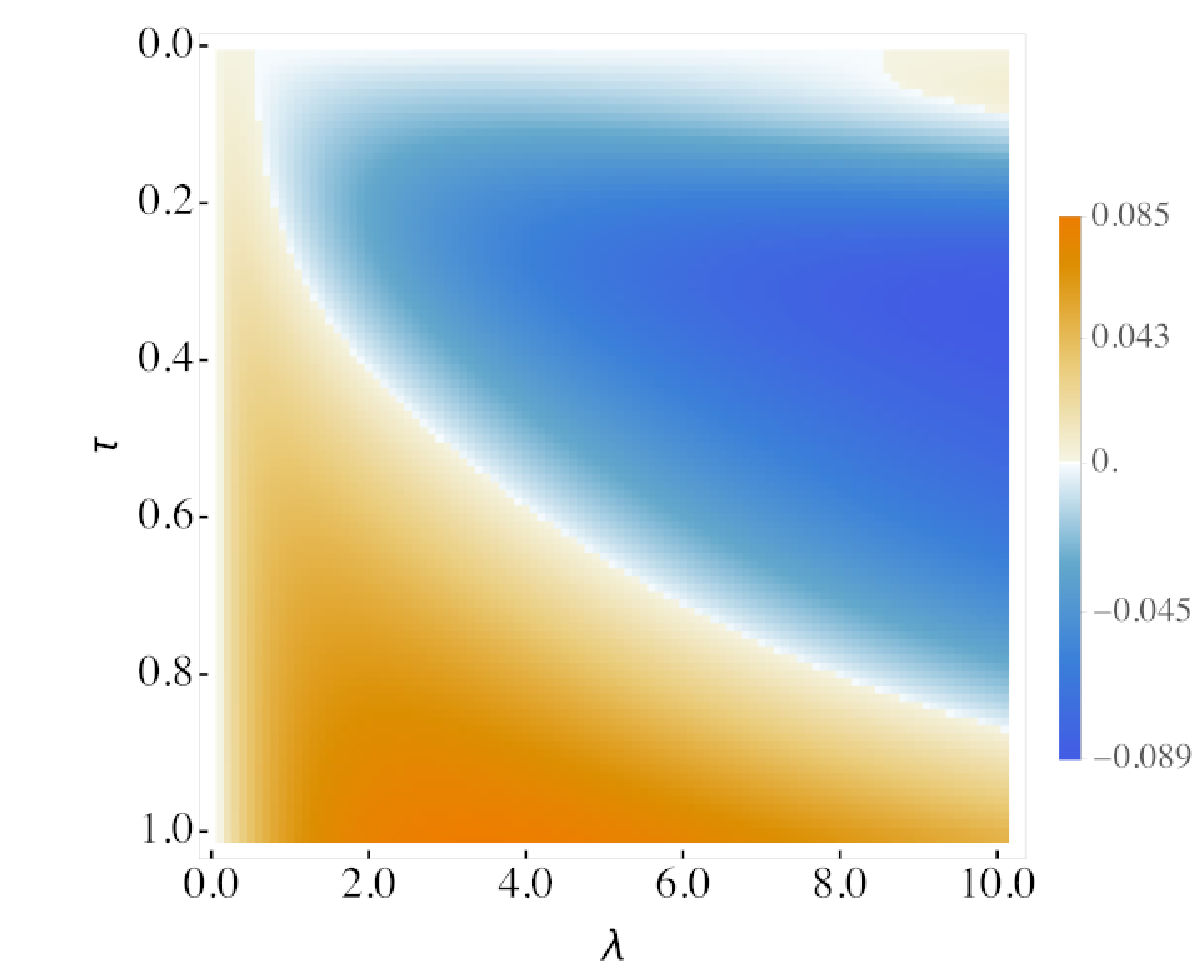}
            \caption{The temperature derivative of the potential exchange-correlation obtained from the GDSM17 parameterization\cite{GDSM17} is taken beyond physical interaction strength $(\lambda=1)$ and shown for $r\s=1$, $0\leq\tau\leq 1$ and $0\leq\lambda\leq10$.}
            \label{fig:beyond1G}
       \end{figure}

The temperature derivative of the potential exchange-correlation obtained from the GDSM17 parameterization is investigated for interaction strengths beyond the physical interaction strength $(\lambda=1)$ and shown in Fig. \ref{fig:beyond1G} for $r\s=1$ and $0\leq\tau\leq1$. Taking $\lambda$ beyond $1$ reveals that for $\tau \approx 0$, the temperature derivative of the exchange-correlation potential eventually becomes positive again in the $\lambda \approx 8.6$ region. The temperature derivative in Fig. \ref{fig:beyond1G} reaches a minimum value of $-0.089$ and a maximum value of $0.085$.

The temperature derivative of the potential exchange-correlation obtained from the corrected KSDT14 and GDSM17 parameterizations was then used to calculate the exchange-correlation entropy via (\ref{eq:Sxc}). The exchange-correlation entropy calculated at reduced temperatures between zero and unity is shown in Fig. \ref{fig:Sxc3} for three different densities. 
       \begin{figure}
            \centering
            \includegraphics[width=\columnwidth]{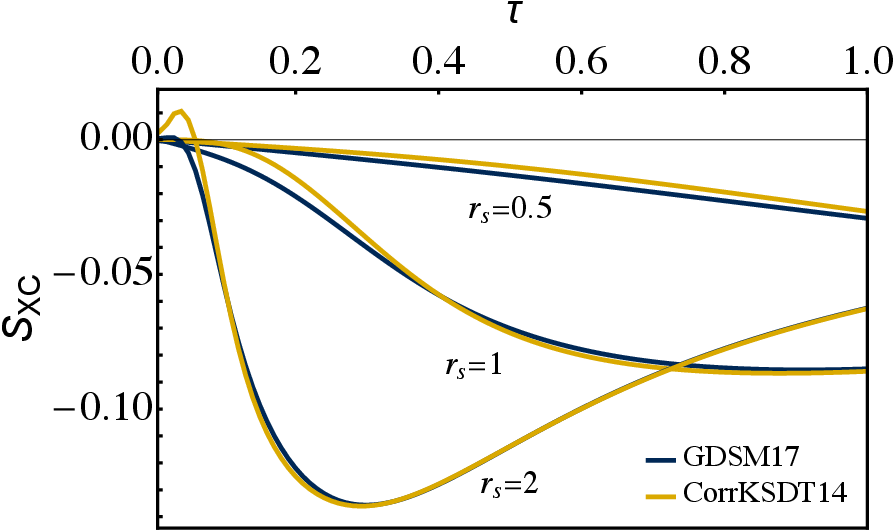}
            \caption{The exchange-correlation entropy is shown for the GDSM17 and corrected KSDT14 paramaterizations of the exchange-correlation free energy of the UEG at WDM conditions at $r\s=0.5,1,2$ and $\tau\rightarrow1$. The behavior exhibited by the $r\s=2$ curves is mimicked by the other two $r\s$ values, but at higher $\tau$.}
            \label{fig:Sxc3}
       \end{figure} 
The exchange-correlation entropy values extracted using the corrected KSDT14 parameterization agree within $0.015$ Ha of the corresponding values obtained from the GDSM17 parameterization, as demonstrated by the difference plot in Fig. \ref{fig:Error}.

       \begin{figure}[htb]
            \centering
            \includegraphics[width=\columnwidth]{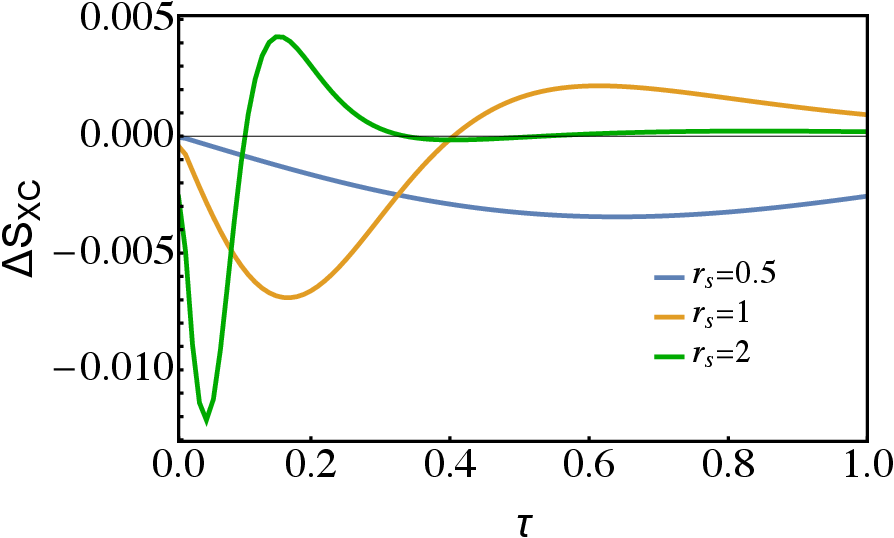}
            \caption{The exchange-correlation entropy calculated using the GDSM17 parameterization is compared to the exchange-correlation entropy calculated from the corrected KSDT14 parameterization and the difference is plotted for $r\s=0.5,1,2$ as $\tau\rightarrow1$. The small discrepancies in exchange-correlation entropy values between the two parameterizations appear to oscillate in sign, indicating crossover of the two parameterizations, are larger at smaller values of $\tau$, then diminish as $\tau$ increases.}
            \label{fig:Error}
       \end{figure}

What is not readily evident in Fig. \ref{fig:Sxc3} is that at  high-enough reduced temperatures, the $r\s=0.5$ and $r\s=1$ curves eventually mimic the qualitative behavior exhibited by the $r\s=2$ curves for $0\leq\tau\leq1$. For the lower values of $r\s$, which correspond to higher densities, the largest magnitude of the exchange-correlation entropy is reached at higher temperatures. For example, the exchange-correlation entropy curves plotted for $r\s=1$ in Fig. \ref{fig:Sxcrs1t5} reach a maximum magnitude at $\tau\approx1$, while the curves plotted for $r\s=0.5$ in Fig. \ref{fig:Sxcrspt5t10} reach a maximum at $\tau\approx3$. Maximum XC entropy values obtained from the GDSM17 and corrected KSDT14 parameterizations are compared in Table \ref{table:maxSxc}.
       \begin{figure}[htp]
            \centering
            \includegraphics[width=\columnwidth]{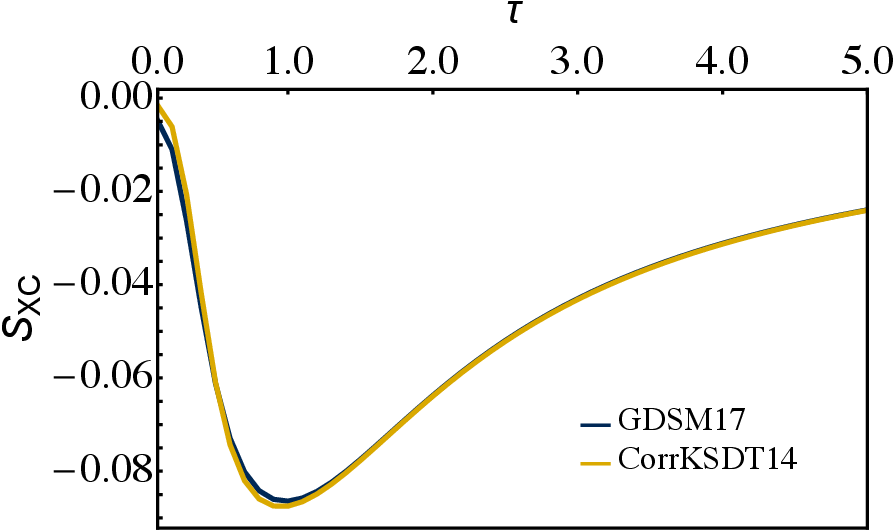}
            \caption{The exchange-correlation entropy for $r\s=1$ and $\tau\rightarrow5$ is compared for the corrected KSDT14 and GDSM17 parameterizations.}
            \label{fig:Sxcrs1t5}
       \end{figure}       
       
       \begin{figure}[htp]
            \centering
            \includegraphics[width=\columnwidth]{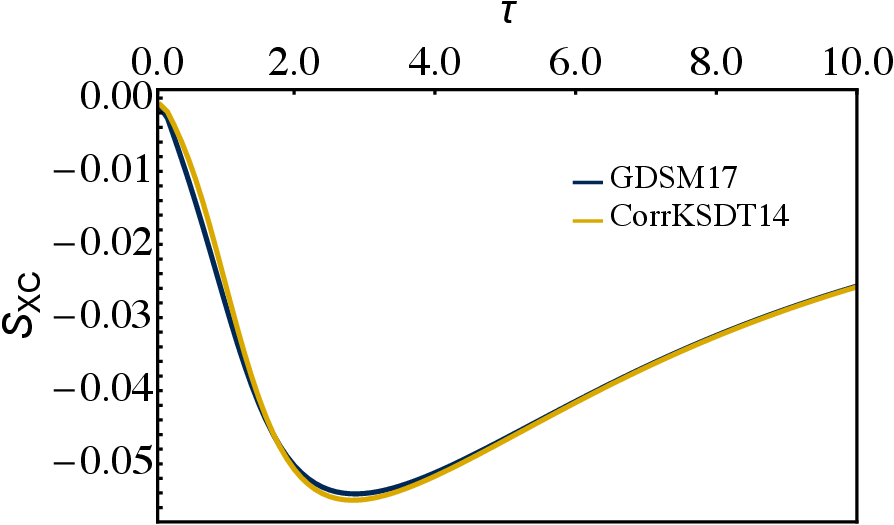}
            \caption{The exchange-correlation entropy for $r\s=0.5$ and $\tau\rightarrow10$ is compared for the corrected KSDT14 and GDSM17 parameterizations.}
            \label{fig:Sxcrspt5t10}
       \end{figure}       
The differences between curves obtained from the corrected KSDT14 parameterization and the corresponding curves obtained from the GDSM17 parameterization are not artifacts of scaling, but are due to differences in the parameterizations themselves, as demonstrated by Fig. \ref{fig:dAxc}. These differences are muted in practice because of the temperature scaling of entropy in the FT electronic Hamiltonian and the small overall value of the exchange-correlation free energy.



\begin{table}
\begin{tabular}{@{  }l@{  }|@{  } c @{  }c @{  }|@{  } c @{  } c @{  }}
\toprule
\textbf{$r_s$}  & \textbf{GDSM17} & \textbf{$\tau$} & \textbf{CorrKSDT14} & \textbf{$\tau$}\\ \midrule 
0.5        & -0.0535   & 2.82     & -0.0544        & 2.78  \\
1.0        & -0.0856   & 0.90     & -0.0867        & 0.87  \\
2.0        & -0.1357   & 0.29     & -0.1361        & 0.29  \\
4.0        & -0.2124   & 0.09     & -0.2108        & 0.09  \\
\bottomrule
\end{tabular}
\caption{The maximum magnitude of the exchange-correlation entropy calculated at three densities is shown for two parameterizations, GDSM17 and the corrected KSDT14, along with the temperature $\tau$ at which each maximum occurs. The maximum XC entropy is converged with $\tau$ and $\lambda$ resolution to within 0.000X.}
\label{table:maxSxc}
\end{table}

\sec{Conclusions}

We have defined a generalized thermal adiabatic connection that depends on a fictitious temperature as well as the coupling constant, while maintaining the scaling relationships we know to be true for the true temperature, coupling constant, and coordinate system. By having $V\ee$, and therefore $A\t\xc$, scale with both coupling constant $\lambda$ and fictitious temperature $\tilde{\tau}$, we are in effect scaling between two temperature-referenced systems (that of a finite-temperature reference system and that at the true temperature $\tau$). Since this is accomplished via an ACF-type process, there are links to prior work in thermal stitching\cite{SB18} we have yet to fully explore. This work used a uniform gas $a\xc$ parameterization for a proof of principle in a system with no variation in density, even implicitly with temperature, so further work should include analysis of more complicated model systems, such as the slowly varying gas. 

In analyzing the conditions at which the extreme values of $S\xc$ are achieved in Table \ref{table:maxSxc}, it appears that the maximum magnitudes occur at pairs of $r\s$ and $\tau$ loosely following, for $n=[0,1,2]$,
\ben
r\s=c 3^{3-n},
\een
and
\ben
\tau=2^{n-1},
\een
with $c$ a small, positive constant. Since these points occur at minimum values, this simple pattern may provide a simple form for approximating the gradient of $S\xc$ as a function of $\lambda$ and $\tau$. Further testing is needed to see if this pattern holds beyond these few test points, as well as if they hold beyond the uniform gas system being tested here. Equivalent scaling with $n$ is achieved when $n=2$, suggesting through the familiar tied temperature-coordinate-interaction strength scaling relations that this form might be used as a sort of exchange-correlation enhancement factor that preserves the scaling of the exchange-correlation free energy per particle of the uniform gas underlying certain approximate forms.

These initial investigations with GTAC also show that the variation of the FT ACF integrand with temperature shifts in character as we move to strong interaction. The regime characterized by strong interaction, low temperature, and intermediate density shows temperature dependence resembling that of low-temperature, weak-interaction regimes, which may hint at dominance of strong interaction effects over thermal effects. This is likely tempered by certain high-density effects, but further tests are necessary to be sure. Surely we will also see differences in non-uniform systems as well, inviting comparison studies for the finite-temperature asymmetric Hubbard dimer and other model systems. The current work points toward a ``sweet spot" of strong effects at moderate values, where temperature, interaction strength, and density influences are all at play. This is similar to what is seen in prior work on thermal DFT\cite{SPB16} and zero-temperature DFT.\cite{GP23}

In addition to this pathway of inquiry, work currently in progress includes extracting consistent and approximate $U\xc$ approximations using GTAC and applying GTAC analysis to generalized gradient approximations. Next steps involve applying GTAC within extended parameter regimes, such as more of those beyond the physical interaction strength, and testing generated $S\xc$ approximations in simulations of more realistically complex simulated systems is soon to follow.

The data that support the findings of this study and the notebooks used to produce them are available from the corresponding author upon request.

\sec{Acknowledgments} Fruitful discussions with Vincent Martinetto, Dr. Sara Giarrusso, and Dr. Juri Grossi are gratefully acknowledged. This work is supported by the U.S. Department of Energy, National Nuclear Security Administration, Minority Serving Institution Partnership Program, under Awards DE-NA0003866 and DE-NA0003984. We acknowledge all indigenous peoples local to the site of University of California, Merced, including
the Yokuts and Miwuk, and thank them for allowing us
to live, work and learn on their traditional homeland (see
\url{https://www.hypugaea.com/acknowledgments}).

\section*{References}
\bibliographystyle{unsrt}
\bibliography{thermal,WDM,extras}

\end{document}

%% file: macros.tex

\def\bea{\begin{eqnarray}}
\def\eea{\end{eqnarray}}
\def\ben{\begin{equation}}
\def\een{\end{equation}}
\def\benu{\begin{enumerate}}
\def\enu{\end{enumerate}}

\def\bei{\begin{itemize}}
\def\eei{\end{itemize}}
\def\beit{\begin{itemize}}
\def\eit{\end{itemize}}
\def\benu{\begin{enumerate}}
\def\enu{\end{enumerate}}

\def\n{n}
\def\np{{n^{\prime}}}
\def\npp{{n^{\prime \prime}}}

\def\sss{\scriptscriptstyle\rm}

\def\g{_\gamma}

\def\l{^\lambda}
\def\lfc{^{\lambda=1}}
\def\lo{^{\lambda=0}}

\def\marnote#1{\marginpar{\tiny #1}}
\def\rsav{\langle r_s \rangle}
\def\invdif{\frac{1}{|\br_1 - \br_2|}}

\def\hatT{{\hat T}}
\def\hatV{{\hat V}}
\def\hatH{{\hat H}}
\def\1var{(\bx_1...\bx\N)}

\def\half{\frac{1}{2}}
\def\quart{\frac{1}{4}}

\def\bp{{\bf p}}
\def\br{{\bf r}}
\def\bR{{\bf R}}
\def\bu{{\bf u}}
\def\bx{{x}}
\def\by{{y}}
\def\ba{{\bf a}}
\def\bq{{\bf q}}
\def\bj{{\bf j}}
\def\bX{{\bf X}}
\def\bF{{\bf F}}
\def\bchi{{\bf \chi}}
\def\bof{{\bf f}}

\def\cA{{\cal A}}
\def\cB{{\cal B}}

\def\x{_{\sss X}}
\def\c{_{\sss C}}
\def\s{_{\sss S}}
\def\xc{_{\sss XC}}
\def\Hx{_{\sss HX}}
\def\Hxc{_{\sss HXC}}
\def\xj{_{{\sss X},j}}
\def\xcj{_{{\sss XC},j}}
\def\N{_{\sss N}}
\def\H{_{\sss H}}

\def\t{^{\tau}}

\def\ext{_{\rm ext}}
\def\pot{^{\rm pot}}
\def\hyb{^{\rm hyb}}
\def\HF{^{\rm HF}}
\def\hah{^{1/2\& 1/2}}
\def\loc{^{\rm loc}}
\def\LSD{^{\rm LSD}}
\def\LDA{^{\rm LDA}}
\def\GEA{^{\rm GEA}}
\def\GGA{^{\rm GGA}}
\def\SPL{^{\rm SPL}}
\def\sce{^{\rm SCE}}
\def\PBE{^{\rm PBE}}
\def\DFA{^{\rm DFA}}
\def\TF{^{\rm TF}}
\def\VW{^{\rm VW}}
\def\helm{^{\rm unamb}}
\def\una{^{\rm unamb}}
\def\ion{^{\rm ion}}
\def\HOMO{^{\rm HOMO}}
\def\LUMO{^{\rm LUMO}}
\def\gs{^{\rm gs}}
\def\dyn{^{\rm dyn}}
\def\adia{^{\rm adia}}
\def\I{^{\rm I}}
\def\pot{^{\rm pot}}
\def\sav{^{\rm sph. av.}}
\def\syv{^{\rm sys. av.}}
\def\pnav{^{\rm sym}}
\def\av#1{\langle #1 \rangle}
\def\unif{^{\rm unif}}
\def\LSD{^{\rm LSD}}
\def\ee{_{\rm ee}}
\def\vir{^{\rm vir}}
\def\ALDA{^{\rm ALDA}}
\def\PGG{^{\rm PGG}}
\def\GK{^{\rm GK}}
\def\atom{^{\rm atmiz}}
\def\trans{^{\rm trans}}
\def\unpol{^{\rm unpol}}
\def\pol{^{\rm pol}}
\def\sav{^{\rm sph. av.}}
\def\syv{^{\rm sys. av.}}

\def\up{_\uparrow}
\def\dn{_\downarrow}
\def\upp{\uparrow}
\def\dnn{\downarrow}

\def\shalf{$^{\hspace{0.006cm}1}\! \! \hspace{0.045cm} / _{\! 2}$}

\def\td{time-dependent~}
\def\KS{Kohn-Sham~}
\def\DFT{density functional theory~}

\def\fourint{ \int_{t_0}^{t_1} \! dt \int \! d^3r\ }
\def\fourintp{ \int_{t_0}^{t_1} \! dt' \int \! d^3r'\ }
\def\intx{\int\!d^4x}
\def\sph_int{ {\int d^3 r}}
\def\radint{ \int_0^\infty dr\ 4\pi r^2\ }
\def\intrrp{\int d^3r \int d^3r'\,}
\def\intr{\int d^3r\,}
\def\intrp{\int d^3r'\,}